\documentclass[
aip, pop,
 amsmath,amssymb,
 reprint,%
tightenlines,
flushbottom,
citeautoscript,
floatfix
]{revtex4-2}

\usepackage{graphicx}% Include figure files
\usepackage{dcolumn}% Align table columns on decimal point
\usepackage{bm}% bold math
\usepackage{comment}
\usepackage{booktabs}
\usepackage{wasysym}
\usepackage[separate-uncertainty=true]{siunitx}
\usepackage[utf8]{inputenc}
\usepackage[T1]{fontenc}
\usepackage{mathptmx}
\usepackage{etoolbox}
\usepackage{hyperref}
\hypersetup{
    colorlinks = true,
    linkcolor = blue,
    citecolor = blue,
    urlcolor=blue,
}
\usepackage{tabularx}
\usepackage{xcolor}
\usepackage{xspace}

\makeatletter
\def\@email#1#2{%
 \endgroup
 \patchcmd{\titleblock@produce}
  {\frontmatter@RRAPformat}
  {\frontmatter@RRAPformat{\produce@RRAP{*#1\href{mailto:#2}{#2}}}\frontmatter@RRAPformat}
  {}{}
}%
\makeatother

\usepackage{setspace}
\usepackage{parskip}
\usepackage[compact]{titlesec}         % you need this package
\titlespacing{\section}{4pt}{4pt}{4pt} % this reduces space between (sub)sections to 0pt, for example
\AtBeginDocument{%                     % this will reduce spaces between parts (above and below) of texts within a (sub)section to 0pt, for example - like between an 'eqnarray' and text
  \setlength\abovedisplayskip{4pt}
  \setlength\belowdisplayskip{4pt}
  }

\usepackage{natmove}

\begin{document}

\title{Delayed current sheet formation due to an external field in pulsed-power-driven reconnection experiments}
% \title{Delayed magnetic reconnection onset by an external field in pulsed-power-driven experiments}
%Driven Magnetic Reconnection in an Externally-Applied Magnetic Field
%Experimental evidence for the delay of reconnection onste by an external magnetic field
%Delayed magnetic reconnection onset by an external field in pulsed-power driven experiments

% Experimental evidence for the delay of reconnection onset by an external magnetic field in a driven magnetic reconnection experiment
% ... driven by pulsed power



\author{T. W. O. Varnish}
\affiliation{Plasma Science and Fusion Center, Massachusetts Institute of Technology, Cambridge, MA 02139, USA\looseness=-10000 %\\This line break forced with \textbackslash\textbackslash
}%

\author{G. V. Dowhan}
\altaffiliation[Present address: ]{Naval Research Laboratory, Washington, DC 20375, USA\looseness=-10000}
\affiliation{Department of Nuclear Engineering and Radiological Sciences, University of Michigan, Ann Arbor, MI 48109, USA\looseness=-10000}%

\author{M. Chen}
\affiliation{Laboratory of Plasma Studies, Cornell University, Ithaca, NY 14853, USA\looseness=-10000 %\\This line break forced with \textbackslash\textbackslash
}%

\author{D. M. Johnson}
\affiliation{Laboratory of Plasma Studies, Cornell University, Ithaca, NY 14853, USA\looseness=-10000 %\\This line break forced with \textbackslash\textbackslash
}%

\author{N. M. Jordan}
\affiliation{Department of Nuclear Engineering and Radiological Sciences, University of Michigan, Ann Arbor, MI 48109, USA\looseness=-10000}%

\author{J. Lee}
\affiliation{Laboratory of Plasma Studies, Cornell University, Ithaca, NY 14853, USA\looseness=-10000 %\\This line break forced with \textbackslash\textbackslash
}%

% \author{D. K. Robinson}
% \affiliation{Plasma Science and Fusion Center, Massachusetts Institute of Technology, MA 02139, Cambridge, USA\looseness=-10000 %\\This line break forced with \textbackslash\textbackslash
% }%

\author{A. P. Shah}
\altaffiliation[Present address: ]{University of California San Diego, La Jolla, CA 92093, USA\looseness=-10000}
\affiliation{Department of Nuclear Engineering and Radiological Sciences, University of Michigan, Ann Arbor, MI 48109, USA\looseness=-10000}%

\author{R. Shapovalov}
\altaffiliation[Present address: ]{Laboratory for Laser Energetics, Rochester, NY 14623, USA\looseness=-10000}
\affiliation{Department of Nuclear Engineering and Radiological Sciences, University of Michigan, Ann Arbor, MI 48109, USA\looseness=-10000}%

\author{B. J. Sporer}
\altaffiliation[Present address: ]{TAE Technologies, Inc., Foothill Ranch, CA 92610, USA\looseness=-10000}
\affiliation{Department of Nuclear Engineering and Radiological Sciences, University of Michigan, Ann Arbor, MI 48109, USA\looseness=-10000}%

\author{R. D. McBride}
\affiliation{Department of Nuclear Engineering and Radiological Sciences, University of Michigan, Ann Arbor, MI 48109, USA\looseness=-10000}%

\author{J. D. Hare}
\email[]{jdhare@cornell.edu}
\affiliation{Plasma Science and Fusion Center, Massachusetts Institute of Technology, Cambridge, MA 02139, USA\looseness=-10000 %\\This line break forced with \textbackslash\textbackslash
}%
\affiliation{Laboratory of Plasma Studies, Cornell University, Ithaca, NY 14853, USA\looseness=-10000 %\\This line break forced with \textbackslash\textbackslash
}%

\date{\today}

\begin{abstract}
We present results from pulsed-power-driven magnetic reconnection experiments, in which we drove two exploding wire arrays in parallel to produce colliding plasma flows with anti-parallel magnetic fields of \SI{1.2\pm0.2}{\tesla}.
The experimental volume was surrounded by a Helmholtz coil pair capable of externally applying a field of up to 2 T, parallel to the reconnecting electric field.
We diagnosed these experiments using laser interferometric imaging in the direction of the anti-parallel magnetic fields, gated extreme ultraviolet pinhole imaging, and in situ inductive probes.
For zero and weak (0.5 T) external fields, we reproduce previous observations in which a dense reconnection layer forms between the two wire arrays.
However, when we apply a strong external field (2 T), we observe a void between the arrays rather than a dense layer, and we hypothesise that the external field is frozen out of the plasma and provides a back-pressure which decelerates the flows.
Our experimental results are compared with three-dimensional magnetohydrodynamic simulations of the experiment, which qualitatively support this hypothesis. 
These simulations allow us to study the pressure balance and dynamics of the current sheet aspect ratio, demonstrating the delayed formation of the reconnection layer due to the external field.
% These results offer insight into the trigger problem of magnetic reconnection, as the external field delays the onset of reconnection and allows magnetic energy to be stored.
\end{abstract}

\maketitle 
\section{\label{sec:intro}Introduction}
Magnetic reconnection explosively reconfigures the topology of magnetic field lines, rapidly dissipating magnetic energy by heating and accelerating plasma.\cite{Zweibel2016}
The reconnection process itself has been studied extensively in theory, simulation, and experiment, most often in systems where reconnection is already occurring.
%Interesting questions here involve the rate at which reconnection occurs, the scales on which it occurs, and how the dissipated magnetic energy is partitioned between heating and acceleration of the plasma.
%However, much less is known about how the system transitions from slowly building up magnetic energy to rapidly reconnecting it, and this is known as the onset or the trigger problem.\cite{Uzdensky2016a}
In this paper, we present an experiment in which we study the role of an external magnetic field in delaying the formation of a reconnection layer.

Reconnection experiments with externally applied magnetic fields have been carried out on laser facilities.
Fiksel \textit{et al.} used a pulsed magnetic coil set on the OMEGA laser to produce an X-point magnetic field configuration.\cite{Fiksel2014}
Two lasers were focused onto solid targets to produce counter-propagating plasma plumes, which compressed the externally applied field.
However, reconnection was only observed when the volume between the colliding plumes was pre-filled with a low-density plasma, which froze in the external field.
In this configuration, the authors observed the plasma plumes gradually decelerated before stagnating, followed by a rapid burst of reconnection at close to the local Alfv\'enic rate. 
It was hypothesized that without the pre-fill plasma, the vacuum magnetic fields were free to reorient themselves out of the way of the driving plasma flows.
We emphasize that the magnetic field geometry in Fiksel \textit{et al.} is different from that presented in this paper. 
Fiksel \textit{et al.} had only externally applied fields in an X-point geometry, which permitted reconnection, whereas we have magnetized plasma flows with advected fields in a reconnection geometry, interacting with an externally applied field which is parallel to the reconnecting electric field (Fig. \ref{fig:hh_setup}).

%Laser-plasma experiments to study magnetic reconnection in a plasma with an externally-generated field have previously been conducted by Fiksel \textit{et al.} on the OMEGA laser facility\cite{Fiksel2014}. In these experiments, a low-density pre-fill plasma was threaded with an externally generated magnetic field, forming magnetic ribbons with oppositely directed fields. These ribbons were driven together by colliding laser-plasma plumes, which forced the frozen-out magnetic ribbons together. Fiksel \textit{et al.} observed these inflows gradually decelerated before stagnating, followed by a rapid burst of reconnection, much faster than steady-state rates, and close to the local Alfv\'enic rate. Interestingly, this was seen only with the presence of a pre-fill plasma for the external field to thread through. Without the pre-fill plasma, there was no evidence of reconnection. It was hypothesized that without the pre-fill plasma, the vacuum magnetic fields were free to reorient themselves out of the way of the driving plasma flows in these experiments.

% Previous work (PULSED-POWER)
% Flux compression - Z-machine/COBRA/MAIZE?
% MagLIF concept 
% * Slutz+ 2010 PoP
% * Gomez+ 2014 PRL
% * Sinars+ 2020 PoP
% * Yager-Elorriaga+ 2022 Nucl. Fusion

Pulsed-power-driven experiments have also been carried out with externally applied magnetic fields.
Compression of an externally-applied magnetic field has been studied extensively using gas-puff z-pinches on MA-scale facilities\cite{Felber1988, Gourdain2013, Seyler2020, Aybar2021, Chen2024b, Lavine2024}.
Here, the field is frozen into an initially low-density plasma, which is compressed along with the plasma column by a rapidly rising current pulse.
Magnetic flux compression is also particularly relevant to the MagLIF fusion concept, as it provides the necessary thermal insulation between the hot fuel and the cold imploding liner \cite{Slutz2010a, Gomez2014, Sinars2020, Yager-Elorriaga2022, Gomez2025}. 

%%% ASTROPHYSICAL IMPLICATIONS

Previous pulsed-power-driven reconnection experiments have used two exploding wire-arrays placed side-by-side and driven in parallel.\cite{Hare2018a}
In an exploding wire array, plasma is created on the wire surfaces by Ohmic heating, and accelerated radially outwards by the $\mathbf{J}\times\mathbf{B}$ force into a vacuum region (see Fig. \ref{fig:hh_setup}).
The plasma flows advect a fraction of the driving magnetic field, producing radially diverging magnetised plasma flows.
When the flows from the adjacent wire arrays collide, they have an anti-parallel field component that reconnects ($B_{rec}$), producing a hot, dense reconnection layer at the midplane between the two arrays.

Prior experiments have observed radiative cooling of the layer using aluminium wires \cite{Suttle2016}, and the formation of plasmoids using carbon wires.\cite{Hare2017}
The wire arrays have also been tilted so that the reconnecting magnetic fields are not exactly anti-parallel---this adds a so-called guide field which does not reconnect, but does interact with two-fluid effects to create a quadrupolar density structure.\cite{Varnish2025}
This platform has been scaled from university-scale machines (\SI{1}{\mega\ampere} peak current) to the Z Machine (Sandia National Laboratories, \SI{20}{\mega\ampere} peak current), where both plasmoids and radiatively cooling were observed simultaneously.\cite{Datta2024b}
In all cases, the plasma flows from the wire arrays expand into a vacuum region, which is devoid of any magnetic field.

In this paper, we present the first results from pulsed-power-driven reconnection experiments in which the exploding wire arrays sit inside a pulsed Helmholtz coil, which provides a vacuum magnetic field into which the plasma flows expand.
With zero or weak (\SI{0.5}{\tesla}) external field, we reproduce the observations of previous experiments in which a well-defined reconnection layer forms.
However, when the external magnetic field is strong (\(B_{ext}\approx B_{rec}\)), we observe a void in electron density where we expect the reconnection layer to form. 
% We hypothesise that this external field is frozen out of the plasma flows rather than providing a guide field.
We ascribe this to the ``freezing-out'' of the externally applied magnetic field, as the hydrodynamic timescale is significantly shorter than the diffusion timescale.
Therefore, instead of adding an out-of-plane magnetic field component---a guide field---to the plasma flows from the wire arrays, we hypothesise that the external magnetic field is compressed between the two flows until the magnetic pressure is sufficient to balance the ram pressure of the flows.
% From 3D MHD simulations of our experiments, we see that this magnetic back-pressure decelerates the flows from the two exploding wire arrays and delays the onset of reconnection.
We provide evidence for this interpretation using 3D magnetohydrodynamic (MHD) simulations of the full experimental geometry, which also show that the reconnection process is stalled by high external fields.
By increasing the current in these simulations, we also show that reconnection can occur at later times when some of the external field has diffused into the plasma.
At this time reconnection proceeds with a guide field ratio $b = B_g/B_{rec} = 0.5$.
% The flows then decelerate due to the back-pressure, delaying the onset of reconnection.

% In this paper, we present results from experiments and 3D magnetohydrodynamic (MHD) simulations, in which we study pulsed-power-driven reconnection with a variable-strength external magnetic field.
% With a weak or zero external field, we observe the formation of a dense reconnection layer, but with a strong external field, no reconnection layer is observed.
% We calculate that the external field is frozen out of the plasma and is compressed in the vacuum region between the two converging flows.
% The pressure of this compressed external field is sufficient to stall the ram pressure of the incoming plasma.
% Simulations support this interpretation and also show that reconnection does occur later, when the ram pressure is sufficient to overcome the external field pressure.
% Therefore, a strong external field significantly delays the onset of reconnection.

\section{\label{sec:method}Methods}

\begin{figure}[!ht]
    \includegraphics{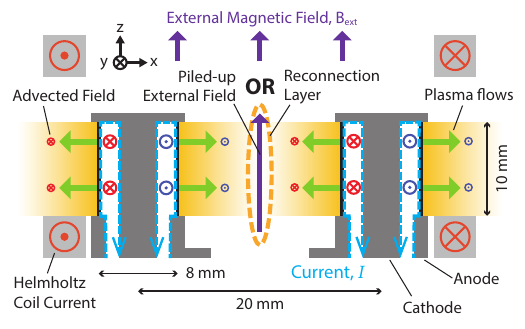}
    \caption{\label{fig:hh_setup} 
    A reconnection layer is formed from the interaction of magnetized outflows from two exploding wire arrays positioned side-by-side.
    An intense current pulse ($\SI{240}{\nano\second}$ rise time, $\SI{400}{\kilo\ampere}$ peak) Ohmically heats the thin (\SI{0.4}{\mm} diameter) carbon rods surrounding a thick (\SI{4}{\mm} diameter) return post.
    The $\mathbf{J} \times \mathbf{B}$ force pushes the coronal plasma around these wires radially outwards, advecting some magnetic field (\SI{1.2\pm0.2}{\tesla}) along with it.
    This entire dual exploding wire array setup was placed inside an externally powered Helmholtz coil, which provides a variable-strength, vertically-oriented, external field of up to \SI{2}{\tesla}.
    An interferometer was set up to observe the region between the two arrays, in the same ``side-on'' field of view as illustrated in this diagram.
    }
\end{figure}

We designed a scaled-down version of the reconnection platform developed for MAGPIE (\SI{1.4}{\mega\ampere} peak current, \SI{240}{\nano\second} rise time),\cite{Hare2018a} suitable for the lower peak current on MAIZE (in this experimental series, \SI{400}{\kilo\ampere} peak current, \SI{240}{\nano\second} rise time) \cite{Gilgenbach2009, McBride2018}.
Fig. \ref{fig:hh_setup} shows this configuration: each wire array was \SI{8}{\mm} in diameter and \SI{10}{\mm} tall, with eight \SI{0.4}{\mm} diameter carbon rods (Staedlar Mars Micro Carbon B) spaced evenly around a \SI{4}{\mm} diameter central cathode.
The arrays had a center-to-center separation of \SI{20}{\mm}, giving a distance of \SI{6}{\mm} from the closest wires to the reconnection layer.
These arrays were placed inside a Helmholtz coil with a \SI{50}{\mm} bore, which was externally energised (\SI{1.2}{\milli\farad}, \SI{1.8}{\kilo\volt} capacitor bank with a \SI{2}{\ms} flat-top discharge)\cite{Campbell2018}.
The long rise time of this coil means that the entire experimental volume was filled with a uniform magnetic field (up to \SI{2}{\tesla}) over the entire experimental timescale.

We diagnosed these experiments using laser interferometry, a four-frame extreme ultraviolet (XUV) camera, and inductive (B-dot) probes.
% Previously 47+/-5 uJ
% Now: 800+/-50 uJ at 5% according to G. Dowhan
The laser interferometry used a \SI{800\pm50}{\micro\joule}, \SI{1064}{\nm}, \SI{2}{\ns} pulse (EKSPLA NL-122), with the beam expanded to a diameter of \SI{25}{\mm}. We used a Mach-Zehnder configuration in which the probe beam passes through the plasma, and a reference beam passes around the vacuum chamber.
These two beams recombine on a beamsplitter and are imaged through the same optics onto a DSLR camera (Canon EOS DIGITAL REBEL XS, $3888\times2592$ pixels) with a long exposure (\SI{1}{\s}---the laser pulse-length sets the actual exposure time).
By tilting the recombining beam-splitter, we introduce a linear phase shift across the vacuum interferogram, to serve as the carrier signal which heterodynes the phase introduced by the plasma. 
We unfold these interferograms using the MAGIC2 code \cite{Swadling2014a, Hare2019} to produce line-integrated electron-density ($\langle n_e L_y\rangle$) maps.
For these experiments, the interferometer was fielded in a side-on configuration, aligned along the $y$-axis, so that the reconnecting magnetic fields point into and out of the imaging plane (as illustrated in Fig. \ref{fig:hh_setup}).

The extreme ultraviolet (XUV) camera consists of a set of four pinholes (\SI{100}{\um}, unfiltered) which image onto four quadrants of a microchannel plate, which are independently triggered to form four gated images\cite{Campbell2018}.
The magnification is 0.92, and the geometric resolution of the camera ($L_{geo}$) is \SI{209}{\um}, but the image is diffraction-limited for photon energies $\leq$\SI{45}{\electronvolt} due to the pinhole size.
The exposure time is \SI{5}{\ns}, with an interframe delay of \SI{15}{\ns}.
The XUV camera is aligned along the $x$-axis of the experiment (Fig. \ref{fig:hh_setup})---perpendicular to the interferometry line-of-sight---and so the view of one wire array is blocked by the wire array in front of it.

The inductive probes (or B-dots) consist of a single loop of wire at the end of a coaxial cable.
The probes are calibrated before the shot, and have a nominal area of \SI{4}{\mm\squared}.
The B-dots are placed inside the bore of the Helmholtz coil, at a distance of \SI{3.5\pm0.5}{\mm} from the wires on the backside of one of the arrays (on the side furthest from the adjacent array, along $x$), and as such are immersed in the plasma.
The radial distance of the B-dot from the wires is verified using measurements taken from pre-shot photographs of the probes, since the radial distance changed shot-to-shot as the probes had to be re-positioned.
The loops are oriented to measure the advected magnetic fields from the array: the same fields which become the reconnecting magnetic field in the layer.
% The inductive ($dB/dt$) signal is retrieved by common mode rejection using the signal from the two counter-wound loops.
The load current is measured using a Rogowski coil in the power feed of MAIZE, which provides a measurement of $dI/dt$.

\section{\label{sec:results}Results}
\begin{figure*}[!hb]
    \includegraphics[width=\textwidth]{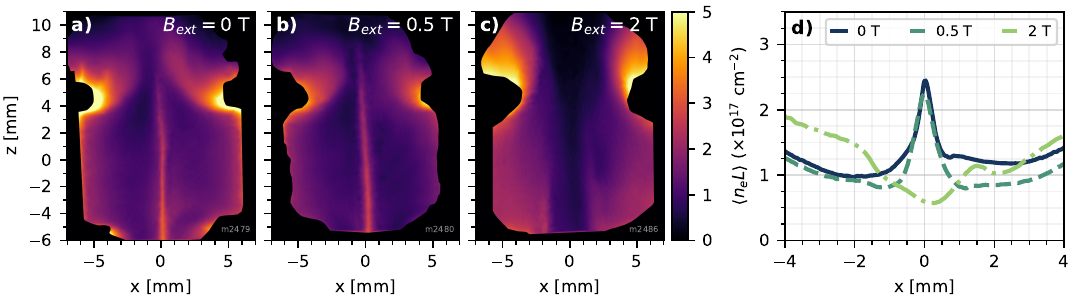}
    \caption{\label{fig:hh_results}\label{fig:if_lineouts} (a--c) Line-integrated electron density maps produced using laser interferometry for a) the zero external field case ($t=\SI{280}{\ns}$), which shows an enhanced region of electron density near $x = \SI{0}{\mm}$, consistent with the formation of a reconnection layer,
    b) the weak external field case, $B_{ext} = \SI{0.5}{\tesla}$ ($t=\SI{270}{\ns}$), where there is still a dense layer between the arrays, and 
    c) the strong external field case $B_{ext} = \SI{2}{\tesla}$ ($t=\SI{280}{\ns}$), where there is a void rather than a dense layer. (d) Horizontal lineouts of the line-integrated electron density maps presented in (a--c), along $z = \SI{0}{\mm}$, averaged over $\Delta z = \pm\SI{2}{\mm}$. Both the \SI{0}{\tesla} and \SI{0.5}{\tesla} lineouts show a peak in line-integrated electron density at the center ($x = \SI{0}{\mm}$), indicative of a reconnection layer. In the \SI{2}{\tesla} lineout, however, a reduction in the density---a void---is seen instead of a layer. We also see a pile-up of the line-integrated electron density in the \SI{2}{\tesla} inflow regions ($|x|\geq\SI{2}{\mm}$).
    }
\end{figure*}

We carried out a series of experiments with this new platform, in which we varied the strength of the external magnetic field between \SIrange{0}{2}{\tesla}.
We summarise our key result in Fig. \ref{fig:hh_results}, which shows three line-integrated electron density maps from one shot with zero external field (Fig. \ref{fig:hh_results}a), one with a weak (\SI{0.5}{\tesla}) external field (Fig. \ref{fig:hh_results}b), and one with a strong external field, $B_{ext} = \SI{2}{\tesla}$ (Fig. \ref{fig:hh_results}c).
All electron density maps are captured at around \SI{280}{\ns} after current start, shortly after peak current at \SI{240}{\ns}.
We measure similar peak currents of \SI{400}{\kilo\ampere} in each of the shots.
In all three line-integrated electron density maps, we can see the exploding wire arrays on the left and right of the image.
The plasma flows are accelerated from these arrays towards the center, advecting with them azimuthal magnetic fields, which point into and out of the plane of the line-integrated electron density map.
We also show lineouts of the line-integrated electron density for the three external field strengths in Fig. \ref{fig:if_lineouts}, which allows for a quantitative comparison.

In the $B_{ext}=\SI{0}{\tesla}$ case, we see an enhanced region of electron density has formed at the mid-plane between the two wire arrays.
In previous experiments on MAGPIE, Faraday rotation imaging showed that this dense layer was the position at which the $B_y$ component of the magnetic field went to zero, consistent with the formation of a reconnection layer (or current sheet).\cite{Hare2017, Hare2017c, Hare2018a}
The lineout in Fig. \ref{fig:if_lineouts} shows that the layer has a peak line-integrated density of $\langle n_e L_y\rangle = \SI{2.5e17}{\per\cm\squared}$, compared to $\langle n_e L_y\rangle = \SI{1e17}{\per\cm\squared}$ just outside the layer.
We also observe regions of plasma moving upwards and inwards from the top wire disks, at $z = \SI{5}{\mm}$, which is likely due to imperfect contact between the electrode and the wires causing additional ablation.

The case with the weak external field, $B_{ext}=\SI{0.5}{\tesla}$, is very similar to the zero external field case.
We still observe the formation of a dense layer, with a similar peak density and a similar density in the inflows.
We also observe a dense plasma region close to contact between the cathode disk and the wires.

In contrast to the previous two cases, in the shot with a strong external field ($B_{ext} \approx B_{rec}$) of $B_{ext}=\SI{2}{\tesla}$, we observe a very different plasma morphology.
Although the density is still high near the wire arrays, indicating ablation and acceleration of plasma flows towards the mid-plane, we see two key differences from the weaker external field cases.
Firstly, plasma formed at the wire disks at $z = \SI{5}{\mm}$ appears to be directed predominantly upwards, rather than upwards and inwards. 
Secondly, and more strikingly, rather than an increase in the plasma density at the mid-plane between the arrays (which would be consistent with a reconnection layer), we see a region of decreased electron density, with a much lower density than that seen in the inflows.
This is clearly shown by the lineout in Fig. \ref{fig:hh_results}c, which shows a much smaller $\langle n_e L_y\rangle = \SI{0.5e17}{\per\cm\squared}$ at $x = \SI{0}{\mm}$ for the $B_{ext}=\SI{2}{\tesla}$ case.
We also observe slightly higher electron densities in the inflow regions ($\langle n_e L_y\rangle = \SI{1.5e17}{\per\cm\squared}$) as compared to the zero external field case (at roughly $x=\pm\SI{2}{\mm}$).
We calculate that the total number of electrons (\(\int\langle n_e L_y\rangle dx dz\)) between the wire arrays in Figs. \ref{fig:hh_results}a and \ref{fig:hh_results}c is the same for all three cases, and so we conclude that there is the same amount of plasma (the ablation rate has not changed), but the plasma has been redirected.

\begin{figure*}[!hbt]
    \includegraphics[width=\textwidth]{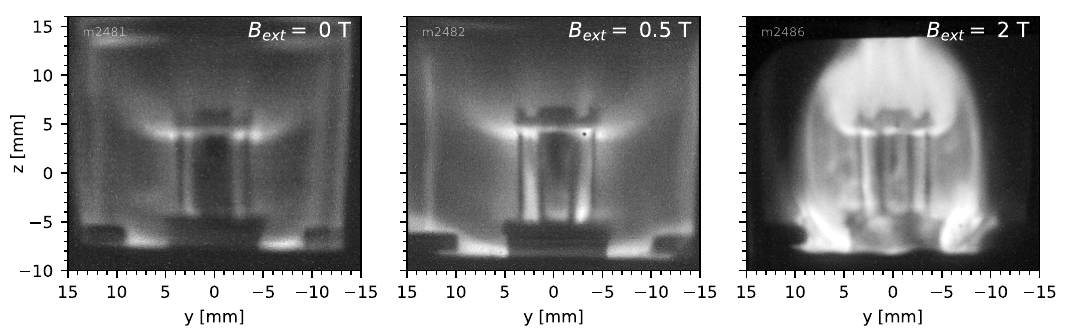}
    \caption{\label{fig:xuv} 
    Images of the extreme ultraviolet (XUV) emission from the plasma, captured using a gated pinhole camera. These images are taken along the $x$-axis, and so one wire array blocks the view of the other wire array. The three images correspond to no external field (left), a weak external field of 0.5 T (center) and a strong external field of 2 T (right). The zero and weak external field cases are qualitatively similar, whereas the strong external field case is noticeably different, with the plasma confined close to the wire array and directed upwards.
    }
\end{figure*}

The redirection of the plasma is particularly evident in the extreme ultraviolet (XUV) self-emission images shown in Fig. \ref{fig:xuv}.
In the zero and weak (0.5 T) external field cases, the plasma emission looks broadly similar, and the plasma fills the field of view as it expands outwards into the vacuum chamber.
Particularly strong emission is seen near the top of the array, as seen in the line-integrated electron density maps, likely due to imperfect wire-electrode contact as mentioned above.
The plasma looks very different in the case of a strong (2 T) external field. 
Here, there is a sudden drop in plasma emission at around y = \SI{\pm10}{\mm} from the center of the array.
There is also a bright region just before the drop in emission, which could be due to a denser or hotter region of plasma, or limb brightening effects.
The XUV emission is also much stronger above the arrays, consistent with the redirection of the plasma flows upwards.

\begin{figure}[!ht]
        \includegraphics[width=\columnwidth]{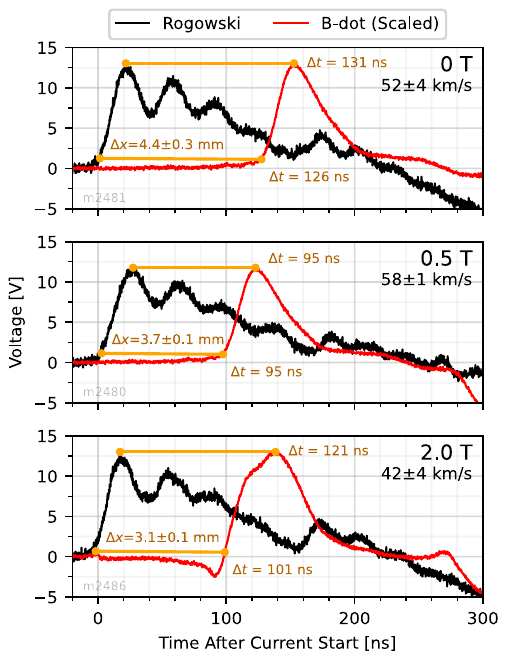}
    \caption{\label{fig:bdot} 
    Comparisons between voltage measurements from the load Rogowski (black) and B-dot probes (red) fielded radially outwards in the outflows of the wire arrays on three separate shots. The B-dot voltage signals are arbitrarily scaled to match the amplitude of the Rogowski signal for easier comparison between the signals in time. There is a time-of-flight effect between the Rogowski signal and the B-dot signal, indicating we are measuring the magnetic field embedded in the flows. The inferred velocities are similar for the zero and weak (0.5 T) external field cases, and lower for the strong (2 T) case.
    }
\end{figure}

Fig. \ref{fig:bdot} presents data from the B-dot probes, which measure the azimuthal magnetic field advected by the plasma in the same three shots as shown in Fig. \ref{fig:hh_results}.
One probe is present in each shot, placed on the backside of an array, and as such, the probe is not in the field of view of the interferometer.
By assuming azimuthal symmetry around the array, we can infer the advected magnetic field and estimate the time-of-flight and hence the flow velocity.
The voltage signal from the Rogowski coil, which measures $dI/dt$ is shown in black, and the signal from the B-dot (scaled to use the same voltage axis) is shown in red.
The time between the start of the Rogwoski signal and the probe signal, and the time between the peaks of the two signals, are both shown in orange.

From this time-of-flight measurement and the known B-dot position (which varied shot-to-shot) we can infer the flow velocity.
For the three shots presented in Fig. \ref{fig:bdot}, we see a similar inferred velocity for the 0 T and 0.5 T external field cases, and a lower velocity for the 2 T case.
We note that in other shots (not shown), these time-of-flight estimates vary, and so we stress that the trend of decreasing flow velocity with increasing external field is weak.
The time-integrated Rogowski and B-dot signals (not shown) provide the current for each shot (well-approximated by $I = \SI{400}{\kilo\ampere} \sin\left[(\pi/2) (t/\SI{240}{\ns})\right]$, and the advected magnetic field which peaks at \SI{1.2\pm0.2}{\tesla} in each shot.

%\begin{figure}[!ht]
%    \includegraphics{figures/bdot_int.pdf}
%    \caption{\label{fig:bdot_int} 
%    Integrated B-dot results...
%    }
%\end{figure}

\section{\label{sec:discussion}Discussion}
Our key observation is that a dense reconnection layer forms for both zero and weak (0.5 T) external fields, but no layer is observed with a strong (2 T) external field.
In addition, the XUV images show that the plasma expansion in the radial direction is limited by a strong external field, and that plasma appears to be redirected upwards.

To explain these observations, we hypothesise that the external fields are frozen out of the plasma flows from the wire arrays.
The frozen-in flux theorem states that the magnetic flux through a surface in an ideal plasma is constant.\cite{Freidberg2007}
This leads to the advection of magnetic fields by a moving plasma, and the increase or decrease of magnetic field strength as the plasma is compressed or expands.
When the plasma is created by Ohmic heating at the wire surface, it is initially dense and has a small cross-sectional area, $A_0$.
As such, the magnetic flux ($\Phi=B_{z,0} A_0$) of the external field frozen into the plasma is small.
Due to the $\mathbf{J}\times\mathbf{B}$ force, this plasma is accelerated radially away from the wires, and the plasma expands into the vacuum region.
As it expands, the area increases, and therefore the advected $B_z$ component of the field decreases due to flux conservation.
This implies that any $B_z$ field in the plasma flows is small, in contrast to the external magnetic field in the vacuum region.
In the region between the two wire arrays, this vacuum field strength actually increases, as the two colliding flows act as compressing, conducting walls.
This gradient in the magnetic field pressure between the vacuum region and the plasma flows acts to decelerate the flows, and delays the formation of the reconnection layer.

This hypothesis requires that the frozen-in (or -out, as it is in this case) condition approximately holds, which implies that the advective term in the magnetic induction equation is larger than the diffusive term.
This is equivalent to stating that the Magnetic Reynolds Number $R_M = V L_{\nabla B}/\bar{\mu}_0$ is greater than unity (where $V$ is the velocity of the moving plasma, $L_{\nabla B}$ is the magnetic field gradient length-scale, and $\bar{\mu}_0 \propto T_e^{-3/2}$ is the magnetic diffusivity).

In these experiments, our only quantitative measurements are of the line-integrated electron density and the magnetic field from the inductive probes.
We can use the more thoroughly diagnosed MAGPIE experiments to estimate inflow velocity, $v_{in}$, and electron temperature $T_e$\cite{Hare2017}.
On MAGPIE, Thomson scattering was used to measure the flow velocity from the carbon wire arrays, $v_{in} = \SI{50}{\km\per\s}$, and this agrees with our time-of-flight estimates from our B-dot probes (Fig. \ref{fig:bdot}).
Thomson scattering was also used to measure the electron temperature, which increased from $T_e\approx\SI{10}{\electronvolt}$ in the inflows to $T_e\approx\SI{100}{\electronvolt}$ within the layer, due to reconnection heating.

\begin{comment}
    We can estimate the reconnecting magnetic field (\(B_{rec}\)) using a scaling argument.
The driving magnetic field at the wires accelerates the plasma flows radially outwards through the \(\mathbf{J}\times\mathbf{B}\) force.
Some fraction of that driving field is advected by the plasma flows, typically around \(10\%\).
We assume this fraction remains the same as we scale the platform between MAGPIE and MAIZE.
The driving magnetic field scales as $B_\theta\propto I/r$, and so we expect an advected magnetic field to scale as \(B_{2, rec} = (I_2/I_1\times r_1/r_2) B_{1,rec}\).
On MAGPIE, $I_1 = \SI{1400}{\kilo\ampere}$, $r_1 = \SI{8}{\mm}$, and Faraday rotation imaging was used to measure $B_{1,rec} \approx \SI{3}{\tesla}$.\cite{Hare2017} 
On MAIZE ($I_2 = \SI{400}{\kilo\ampere}$, $r_2 = \SI{4}{\mm}$), we therefore estimate $B_{2,rec} \approx \SI{1.5}{\tesla}$, again in agreement with our B-dot results.
\end{comment}

Using these values, we can compare the hydrodynamic timescale to the magnetic diffusion timescale, using the estimated values: $L_{\nabla B}=\SI{6}{\mm}$ (wire-to-layer distance), $v_{in}=\SI{50}{\kilo\meter\per\second}$, $\bar{Z}=4$, $\ln \Lambda = 10$, and $T_e = \SI{10}{\electronvolt}$. 
For our experiments, we find that the hydrodynamic timescale is 
% tau_hydro = 120 ns
% tau_eta = 350 ns
\begin{equation}
    \tau_{hydro} \sim \frac{L_{\nabla B}}{v_{in}} \approx \SI{100}{\ns},
\end{equation}
and the magnetic diffusion timescale is
\begin{equation}\label{eqn:diff}
    \tau_{\eta} \sim \frac{L_{\nabla B}^2}{\bar{\eta}} \approx \SI{400}{\ns},
\end{equation}
 using the classical Spitzer-Braginskii plasma resistivity.\cite{Chen2016a}
As $\tau_{hydro}\ll\tau_{\eta}$, the external field does not have sufficient time to diffuse into the plasma flows, and is instead frozen out of the expanding plasma flows.
Such an effect has also been seen in laser-driven experiments, where a laser-driven plasma plume expands into a strong external magnetic field.\cite{Albertazzi2014, Khiar2019}

As the external field is frozen out, we propose that it is compressed by the colliding flows inside the vacuum region at the center of the experiment, providing an increasingly strong magnetic pressure in the process, which stalls the incoming plasma.
% v_A = 40 km/s, v_in = 50 km/s, M_A = 1.25
Using the measurements of the line-integrated electron density and the magnetic fields, we calculate that the inflows are super-Alfv\'enic ($M_A \approx 1.25$).
Therefore, there is insufficient time for the vacuum magnetic field to reorient and move out of the way of the incoming plasma flows.
Instead, the vacuum magnetic field piles up and is compressed by the incoming plasma flows.
We can confirm whether this mechanism is plausible by comparing the magnetic pressure with the ram pressure of the flows, using the known external magnetic field, the inferred mass density from our measurements of $\langle n_e L_y \rangle$ and our estimates of the flow velocity. For the case of a strong (\SI{2}{\tesla}) external field:
% \begin{align}
% Original calculations
%     P_B = \frac{B^2}{2 \mu_0} &\sim \SI{1}{\mega\pascal}, \\
%     P_{ram} = \rho u^2 &\sim \SI{1}{\mega\pascal}, \\
%     P_B &\gtrsim P_{ram}
% \end{align}
% Calculations for Pressures:
% P_B (before compression)
% P_B (0 T) = 0 Pa
% P_B (0.5 T) = 0.1 MPa
% P_B (2 T) = 2 MPa
% P_ram:
%   m = 2e-26 kg
%   L = 20 mm = 2e-2 m (L_y, extent of layer)
%   Z = 4
%   <neL> (0 T) = 1e17 cm^{-2} = 1e21 m^{-2}
%   <neL> (0.5 T) = 1e17 cm^{-2} = 1e21 m^{-2}
%   <neL> (2 T) =   1.5e17 cm^{-2} = 1.5e21 m^{-2}
%   P_ram (0 T) = 3 MPa
%   P_ram (0.5 T) = 3 MPa
%   P_ram (2 T) = 5 MPa
\begin{align}
    P_{B, \ void} = \frac{B^2}{2 \mu_0} &\approx \SI{2}{\mega\pascal}, \\
    P_{ram, \ in} = \rho v_{in}^2 = m(1 + \frac{1}{\bar{Z}})\frac{\langle n_e L_y \rangle}{L_y} v_{in}^2 &\approx \SI{3}{\mega\pascal}
\end{align}
where we assume $m=\SI{2e-26}{\kg}$, $\bar{Z}=4$, $L_y = \SI{20}{\mm}$, and $v_{in}=\SI{50}{\kilo\meter\per\second}$. $\bar{Z}$ and $v_{in}$ are both assumed from the MAGPIE scaling arguments presented earlier, and $L_y$ is reconnection layer length in the y-direction, inferred from the radius of curvature of the field lines.
Assuming $T_{e} \approx \SI{10}{\electronvolt}$, we find the thermal pressure in the inflows ($\SI{5}{\kilo\pascal}$) is negligible compared to the ram and magnetic pressures, and can be ignored in this comparison.
Initially, the magnetic pressure in the void, $P_{B, \ void}$, is similar to the ram pressure in the flows, $P_{ram, \ in}$, and so as the magnetic field in the void is compressed, the back-pressure will decelerate the flows.

For the weak external field (0.5 T), the magnetic pressure is a factor of $(0.5/2)^2 = 1/16$ weaker ($P_{B, \ void} \approx \SI{0.1}{\mega\pascal}$), and so it is dynamically insignificant compared to the ram pressure ($P_{ram, \ in} \approx \SI{3}{\mega\pascal}$).
This agrees with our experimental observations that the weak external field case is very similar to the zero external field case.

\begin{figure*}[!t]
    \includegraphics[width=\textwidth]{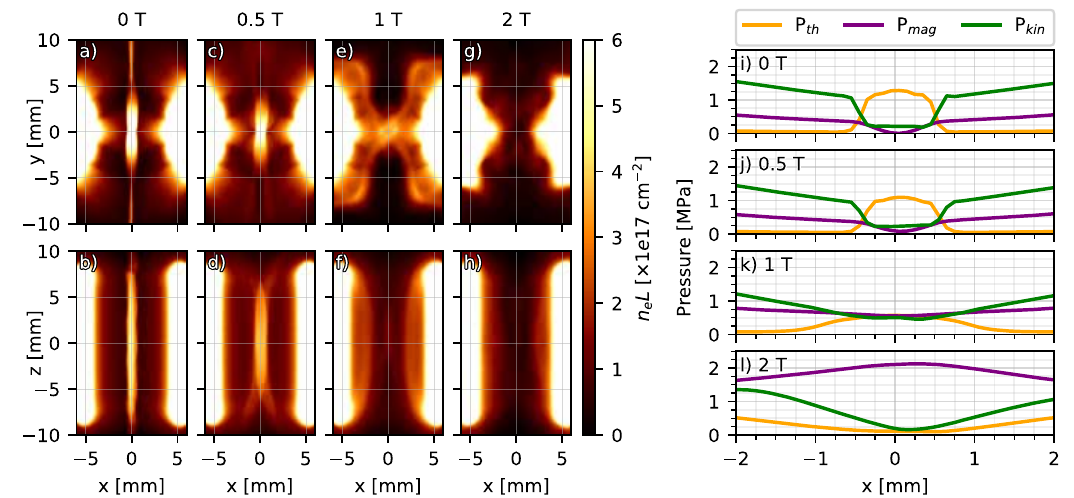} \\\caption{\label{fig:hh_sims_neL}\label{fig:hh_sims_pressure}
    For $I_{peak}=\SI{150}{\kilo\ampere}$: line-integrated electron density maps calculated from three-dimensional MHD simulations at 280 ns after current start. Top row (a, c, e, g): end-on view, bottom row (b, d, f, h): side-on view. First column (a, b): 0 T external field, second column (c, d): 0.5 T external field, third column (e, f): 1 T external field, and fourth column (g, h): 2 T external field.
    These simulations qualitatively reproduce the experimental results shown in Fig. \ref{fig:hh_results}.
    We include results from a 1 T simulation---despite having no experimental data at this external field strength---so we can include it in our later analysis of the layer formation time (Fig. \ref{fig:aspect-ratio}).
    To the right of the line-integrated electron density maps are lineouts of the pressure components (thermal, magnetic, and kinetic) in the plasma across the reconnection layer: i) 0 T external field, j) 0.5 T external field, k) 1 T external field, and l) 2 T external field.
    With no external field (i), the magnetic pressure goes to zero at $x = \SI{0}{\mm}$, but with an external field (j, k, l), the magnetic pressure is non-zero between the arrays, providing a back-pressure which slows layer formation.
    }
\end{figure*}

\section{\label{sec:simulations}Simulations}
To further explore the role of the external magnetic field, we carried out a series of two-dimensional and three-dimensional magneto-hydrodynamic simulations using the GORGON code, an Eulerian resistive MHD code with van Leer advection.\cite{Chittenden2004, Ciardi2007}
We use a \SI{100}{\um} grid with $640\times480\times280$ cells in $(x,y,z)$, and we apply an azimuthal time-varying magnetic field boundary condition at the bottom of each array $(z = \SI{-14}{\mm})$, to deliver half of the total current to each array.
% This total current is split equally between the two arrays, and the magnetic field at the lower $z$ boundary is set by the to be consistent with the current delivered by MAIZE.
The full experimental hardware is modelled, and the external magnetic field is applied as a boundary condition, which fills the simulation domain with an axial magnetic field at $t = 0$.
On the $x$ and $y$ boundaries, we constrained $B_z$ to be continuous for numerical simplicity.
This boundary condition artificially traps the magnetic flux inside the simulation domain, unlike the open vacuum boundaries of the experiment.
However, we increased the domain size until this boundary effect was not significant during the timescales presented below.

We initially used a current pulse of $I = \SI{400}{\kilo\ampere} \sin\left[(\pi/2) (t/\SI{240}{\ns})\right]$, which matched our Rogowski measurements in the experiments.
However, these simulations predicted line-integrated electron densities in the inflow region on the order of $\SI{5e17}{\cm^{-2}}$, which is much denser than seen in our experiments.
%Additionally, it was observed that the reconnection layer formed much earlier in time than would be expected from our experimental results.
Additionally, the magnetic field predicted by the simulations was larger than measured in experiment: at $\SI{4}{\mm}$ radially from the center of the wire cores (at the radial location of the B-dot used experimentally), the azimuthal magnetic field measured in the $\SI{400}{\kilo\ampere}$ simulation was $\SI{4.8}{\tesla}$, compared with $\SI{1.2 \pm 0.2}{\tesla}$ in the experiments.

To address this discrepancy, we ran a series of 2D simulations (in the $x$--$y$ plane at $z=0$) to adjust the peak current, $I_{peak}$, the initial wire temperature $T_{init}$, and the initial expanded wire core diameter $d_{init}$,\cite{Chittenden2004} such that the inflow electron density at $x=\pm\SI{3}{\mm}$ in the simulations matched that seen in our experimental results.
The inflow electron density depends only weakly on $T_{init}$ and $d_{init}$ (which we set to $\SI{0.05}{\electronvolt}$ and $\SI{600}{\um}$ respectively), but it depends strongly on $I_{peak}$.
Reducing the peak current to $\SI{150}{\kilo\ampere}$ gave a synthetic B-dot measurement of $\SI{1.5}{\tesla}$, and a line-integrated inflow electron density of $1 \times 10^{17} \si{\per\cm^{-2}}$, in agreement with experiment.

Although the true reason for this discrepancy in the peak current is not known, we have two working hypotheses.
Firstly, there may be current losses in the load hardware, after the Rogowski coil.
However, there was no evidence for shorting or arcs in any of the other diagnostics.
Secondly, when we simulate wire arrays in MHD codes, we initialise the simulation with the wires replaced by a cold, dense plasma. 
This means we do not track the full evolution of the wires from the solid state to the plasma state, and we do not capture any of the heterogeneity in the exploded wires. 
Properly treating these physical processes likely affects the mass ablation rate from the wires, and hence the electron density we see in the plasma flows.
However, such a detailed treatment is beyond the scope of this paper.
Throughout the remainder of this paper, we will use the \SI{150}{\kilo\ampere} current pulse for the 3D simulations, unless stated otherwise, so that we can qualitatively use the simulations to explore the pressure balance and layer formation.

% \begin{figure}[!ht]
%     \includegraphics[width = \columnwidth]{figures/hh_sims_neL.pdf} \\
%     \caption{\label{fig:hh_sims_neL}
%     Line-integrated electron density maps calculated from three-dimensional MHD simulations at 200 ns after current start. Top row: (a, c, e) end-on view, bottom row: (b, d, f) side-on view. Left column: (a, b) 0 T external field, center column: (c, d) 0.5 T external field, right column: (e, f) 2 T external field.
%     These simulations qualitatively reproduce the experimental results shown in Fig. \ref{fig:hh_results}.
%     }
% \end{figure}

The line-integrated electron densities from these simulations are shown in Fig. \ref{fig:hh_sims_neL}a--h, for the same external field strengths as in the experiments, at the same time after current start ($\SI{280}{\ns}$).
We also include a simulation with an external field of $\SI{1}{\tesla}$, for comparison.
The bottom row shows the side-on line-integrated electron density maps, which should be compared with the experimental results in Fig. \ref{fig:hh_results}.
There is a qualitative match between the simulations and experiments---for the zero and weak external field cases, a dense reconnection layer forms between the two arrays, but for the strong external field case, there is no layer.
In the $\SI{1}{\tesla}$ external field case, we see the early stages of what could be a reconnection layer forming between the two arrays, however it is not as clearly defined as in the zero and weak external field cases.
The electron density maps integrated along the $z$-axis---looking down on the reconnection layer, rather than from a side-on field of view---are shown in the top row. 
Here, we can see that the plasma flows expand into the vacuum region for the zero and weak guide-field cases, but are significantly slowed for the $\SI{1}{\tesla}$ and $\SI{2}{\tesla}$ external field cases.
For the $\SI{1}{\tesla}$ external field, we see two bubble-like regions of plasma emerging from the exploding arrays, with a sudden drop-off in electron density at around $x = \SI{\pm 2.5}{\mm}$.
In the strong ($\SI{2}{\tesla}$) external field case, this drop in electron density is even more sudden, happening closer to the wire arrays, and is consistent with the flows stalling due to the pile-up of frozen-out external magnetic field.

% We note that there are quantitative disagreements between the simulations and the experiments.
% The simulations predict a significantly higher electron density, which may indicate current loss in the experiments.
% Because of this, the ram pressure of the flows in the simulations is higher than estimated for the experiments.
% So, by $t = \SI{280}{\ns}$ the flows have collided even for the strong external field case.
% To make a qualitative comparison, we therefore show simulation results for $t = \SI{200}{\ns}$ after current start, rather than $t = \SI{280}{\ns}$ for the experiments.

% \begin{figure}[!ht]
%     \includegraphics[width = \columnwidth]{figures/hh_sims_pressure.pdf} \\
%     \caption{\label{fig:hh_sims_pressure}
%     lineouts of the pressure components (thermal, magnetic, and kinetic) in the plasma across the reconnection layer. Top: 0 T external field, bottom: 2 T external field.
%     With no external field, the magnetic pressure goes to zero at $x = \SI{0}{\mm}$, but with an external field, the magnetic pressure is non-zero between the arrays, providing a back-pressure which slows layer formation.
%     }
% \end{figure}

We can further probe the role of the piled-up, frozen-out external field by calculating the pressure balance in these simulations, as shown in Fig. \ref{fig:hh_sims_pressure}i--l.
The thermal (orange), magnetic (purple), and kinetic (green) pressures are shown for each external field strength, as a lineout along the $x$-axis (the direction of the inflows).
These lineouts are calculated from pressures averaged over $\SI{\pm0.5}{\mm}$ in $y$, and over $\SI{\pm5}{\mm}$ in $z$.
Only a small region $x = \pm\SI{2}{\mm}$ is shown to provide a detailed look at the dynamics close to the reconnection layer.
Again, the zero and weak external field cases are very similar---kinetic pressure dominates the inflows, but goes nearly to zero at $x = 0$ as the counter-propagating flows collide.
The magnetic pressure also drops to zero in the center of the layer as the reconnection process annihilates the magnetic flux.
The thermal pressure is much higher in the center of the layer due to the increased density and significant heating by the reconnection process, and this thermal pressure supports the layer against the kinetic and magnetic pressures outside.
Although not shown on these plots, the total pressure ($P = P_{th} + P_{mag} + P_{kin}$) remains constant across the layer in all four cases.

In the case of the $\SI{1}{\tesla}$ external field, we see that the thermal, magnetic, and kinetic pressures are all equal at $x = 0$.
This region of equal pressure extends to around $x = \SI{\pm 0.8}{\mm}$, where we see a much smoother transition into the outflow regions than seen in the zero and weak external field cases.
Within this region, the kinetic pressure drops slightly as the flows are decelerated, the thermal pressure rises, and the magnetic pressure stays roughly constant.
At this time ($\SI{280}{\ns}$), we can see the early signs of a reconnection layer, but it is not as clearly defined as in the zero and weak external field cases.

For the strong ($\SI{2}{\tesla}$) external field case, we again see very different results.
The kinetic pressure drops much more rapidly closer to the wires, the magnetic pressure is larger in the inflows and even larger at $x = 0$.
The thermal pressure remains low throughout, though is slightly larger in the inflows compared with the zero and weak ($\SI{0.5}{\tesla}$) external field cases.
This is consistent with the pile-up of the frozen-out external field decelerating the flows, initially preventing the plasma flows from colliding and therefore preventing a reconnection layer from forming.

\begin{figure}[!ht]
        \includegraphics[width=\columnwidth]{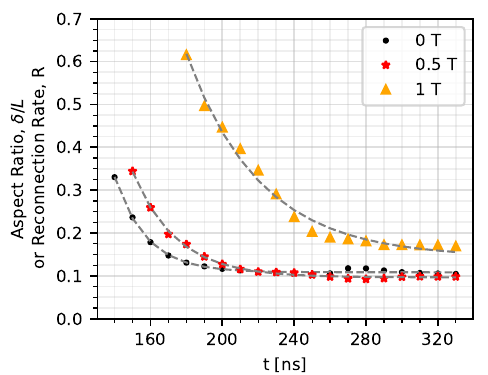}
    \caption{\label{fig:aspect-ratio}
    Aspect ratio ($\delta / L$) of the reconnection layer, plotted for the $\SI{150}{\kA}$ simulations. Measurements of $\delta$ and $L$ are taken from the $z$-directed current density, $J_z$, and shown for the zero external field (black circles), weak ($\SI{0.5}{\tesla}$) external field (red stars), and $\SI{1}{\tesla}$ external field (orange triangles) simulations. Exponential fits to these data are also shown (grey dashed line). As the external field strength is increased, the time at which the layer forms is increased. Data are not shown for the strong ($\SI{2}{\tesla}$) external field simulation, as no reconnection layer was seen.
    }
\end{figure}

Using the $\SI{150}{\kA}$ simulations, we can also study the time evolution of the layer formation.
We take lineouts of the z-directed current density, $J_z$, across the layer, and average them over $\pm\SI{1}{\mm}$ in the $z$ direction.
To measure the layer width, $2\delta$, we take a lineout in the $x$ direction, averaged across $\pm\SI{3}{\mm}$ in $y$, and fit a Gaussian profile to measure the FWHM of the current sheet.
To measure the layer length, $2L$, we take a lineout in the $y$ direction, averaged across $\pm\SI{1}{\mm}$ in $x$, and calculate the FWHM directly.
Fig. \ref{fig:aspect-ratio} shows the layer aspect ratio, $\delta / L$, as a function of time for the zero, weak ($\SI{0.5}{\tesla}$), and $\SI{1}{\tesla}$ external field simulations.
Data from the strong ($\SI{2}{\tesla}$) external field simulation are not shown here, as no current sheet formed during the simulation.
For each of the three external field strengths shown, we see a similar trend: after the initial formation of the current sheet (when we can measure both $\delta$ and $L$), the aspect ratio begins to decrease exponentially, asymptoting to a steady value at late times.
In both the zero and weak ($\SI{0.5}{\tesla}$) external field cases, the layer aspect ratio decreases to $\delta/L \sim 0.1$ on the order of an Alfv\'en time.
With no external field, we observe the initial formation of a current sheet at $\SI{140}{\ns}$.
When the external field is increased to $\SI{0.5}{\tesla}$, we see this formation is delayed by $\SI{10}{\ns}$, forming at $\SI{150}{\ns}$ instead.
However, when the external field is increased further still to $\SI{1}{\tesla}$, we do not observe a layer forming until 180 ns.
Adding a guide field has delayed the formation of the layer.
However, a delay on the order of $\SI{10}{\ns}$ (for the weak external field case) is within the jitter uncertainty of the experiment. 
Due to slight variations in the triggering time or peak current of the pulsed-power facility, repeating the zero external field case would likely result in measuring a slightly different layer formation time for each shot.
It would therefore be difficult to measure this small delay experimentally.
However, for the $\SI{1}{\tesla}$ external field case, our simulation predicts a delay in the formation of the layer of $\sim\SI{40}{\ns}$, which should be much larger than the shot-to-shot jitter.
In future experimental campaigns, we would recommend using a $\SI{1}{\tesla}$ external field to test this hypothesis.

Additionally, the aspect ratio, $\delta / L$, is proportional to the Lundquist number, $S^{-1/2}$.
For an aspect ratio of $\delta / L \sim 0.1$, we therefore measure a Lundquist number, $S \sim 100$.
This is consistent with measurements made on the MAGPIE facility \cite{Hare2017, Hare2018a}.
An exponential decrease in the layer aspect ratio is also expected for a layer described by the Chapman-Kendall X-point collapse model \cite{Chapman1963, Biskamp2000}.
This model is designed for a steady-state, incompressible layer, however our system is compressible and evolving in time.
Additionally, it should be noted that we measure $\delta$ to be on the order of the ion skin depth, $d_i$, throughout the simulations.
As such, two-fluid effects may be important to the dynamics of the layer formation process in our setup.
This is not captured in our simulations, however, as resistive MHD models alone are unable to capture two-fluid effects.
A full analysis of the X-point collapse process observed here is beyond the scope of this paper.

% OUTFLOW REGIONS
% y: 5-15 mm, x: -2, 2 mm, z: -10, 10 mm
% * 2.1e16 in 0 T, 
% * 2.7e16 in 0.5 T
% * 0.8e16 in 2 T

\begin{figure}[!ht]
    \includegraphics[width=\columnwidth]{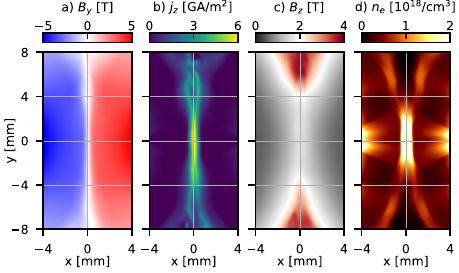}
    \caption{\label{fig:hh_sims_400kA_280}
    For $I_{peak}=\SI{400}{\kilo\ampere}$, we show slices at $z=0$ of components of the magnetic field and current density from 3D MHD simulations at \SI{280}{\ns} after current start. (a) $y$-component of the magnetic field, showing annihilation of the fields at the mid-plane ($x=\SI{0}{\mm}$). (b) $z$-component of the current density, showing a strong current at the mid-plane, indicative of a reconnection layer. (c) $z$-component of the magnetic field, showing a reduction in $B_z$ where the frozen-out flows expelled the field around the arrays ($\pm\SI{4}{\mm}$), a region of $B_z=\SI{2}{\tesla}$ around the reconnection layer, and a compressed field in the outflows ($B_z>\SI{2}{\tesla}$).
    }
\end{figure}

Although we do not have any experimental observations later in the evolution of the plasma, in simulations at $\SI{400}{kA}$ peak current, we observe that eventually the plasma flows do collide and form a reconnection layer, even with a strong ($\SI{2}{\tesla}$) external field.
Fig. \ref{fig:hh_sims_400kA_280} shows maps of the reconnecting magnetic field, $B_y$, the current density, $j_z$, and the strong ($\SI{2}{\tesla}$) externally applied field, $B_z$, at \SI{280}{\ns}.
In these higher-current simulations, an already-established reconnection layer is observed at $\SI{200}{\ns}$.
We see a reversal and annihilation of the $B_y$ field, and a strong $z$-directed current at the mid-plane ($x=\SI{0}{\mm}$), indicative of a reconnection layer.
Importantly, we see that there is a non-zero $B_z = \SI{1.5}{\tesla}$ within the reconnection layer itself, indicating guide field reconnection.
The distribution of $B_z$ within the layer is broad, consistent with a diffusive process.
As the external field is compressed between the counter-propagating plasma flows, the gradient length scale of the field becomes shorter, which in turn reduces the magnetic diffusion timescale, $\tau_\eta$ (Eq. \ref{eqn:diff}).
With $L=\SI{1}{\mm}$, we estimate a diffusion timescale on the order of $\tau_\eta\approx\SI{10}{\ns}$.
This is now a tenth of the hydrodynamic timescale and sufficiently fast for the field to diffuse into the plasma flows.
There is still significant piled-up frozen-out external field downstream of the reconnection layer in the outflows (red regions in Fig. \ref{fig:hh_sims_400kA_280}c), which may provide a back-pressure and slow the reconnection rate.\cite{Ji1999}
Although reconnection does eventually occur, the external field initially prevents layer formation.
Experimental verification of these simulation predictions, including an experimental measurement of the time at which the reconnection layer forms in different external field strengths, will be the subject of a future publication.

% \begin{figure}[!ht]
%         \includegraphics{figures/hh_sims_fields_lo_280ns_cropped.pdf}
%     \caption{\label{fig:hh_sims_280_lo} 
%     Lineouts of the $z$-slices of $B_y$, $j_z / 2$, and $B_z$ presented in Fig. \ref{fig:hh_sims_280}, along $y=0$. Crossing $x=0$, we see a reversal in the $B_y$ value, aligned with a sharp rise in $j_z$, indicative of a reconnection layer. The region where $B_z > 0$ spans a larger width in $x$ ($2\delta\approx\SI{4}{\mm}$) than the reconnection layer (from $j_z$ and $B_y$, which measure a width, $2\delta\approx\SI{1}{\mm}$), suggesting the external field has now diffused into the plasma, particularly across the reconnection layer. Within the layer, the guide field ratio is $b=B_z/B_y\approx0.5$.
%     }
% \end{figure}

\section{\label{sec:conclusions}Conclusions and Future Work}

In this paper, we have presented evidence for the deceleration of magnetized plasma flows as they propagate into a vacuum region with a strong magnetic field.
In the experimental data, we see a void in the line-integrated electron density maps instead of a dense layer when a strong (\SI{2}{\tesla}) external field is applied.
% In 3D MHD simulations of this experiment, we see that this deceleration of the magnetized plasma flows instead delays the formation of the layer rather than completely preventing it.
Using our diagnostics and estimates from similar experiments, we showed that the plasma is initially too conductive to allow the vacuum field to diffuse into the plasma flows, and as such, it was frozen out and piled up in the vacuum region between the flows.
Simulations agree with this hypothesis and show that the magnetic pressure of the piled-up field is comparable to the ram and thermal pressures of the counter-propagating flows, thereby delaying the layer formation.

% \textcolor{red}{Talk about layer formation time here + measurements from simulations.}
From our simulations, we also measured the layer aspect ratio, $\delta/L$, and used this to study the time evolution of the layer formation.
As the external field strength was increased, we observed that the formation of the layer was increasingly delayed.
The layer aspect ratio was also observed to follow an exponential decrease in time, settling at around $\delta/L \sim 0.1$ at late times for both the $\SI{0}{\tesla}$ and $\SI{0.5}{\tesla}$ external fields.

In simulations with a higher driving current than that used in our experiments, we observe that the external field is compressed into a region only \SI{2}{\mm} across, at late times.
At this point, the gradient length scale for the external field is now very short, resulting in very rapid diffusion.
This enables the plasma flows to collide, and for reconnection to occur, with the external field now providing a guide-field component of comparable magnitude to the reconnecting field ($b = 0.5$).

The diagnostics in these experiments were limited to side-on laser interferometry, extreme ultraviolet (XUV) pinhole imaging, and inductive (B-dot) probes.
% From our simulations, we expect very strong heating when the counter-propagating flows are finally dense enough to overcome the back-pressure from the piled-up magnetic field, because of the stored magnetic energy in the anti-parallel fields.
% This heating could also be diagnosed using optical Thomson scattering.
Future experiments could accurately measure the flow velocity using optical Thomson scattering, which could also be used to measure the heating within the reconnection layer.

These experiments were originally designed to embed a guide field into the plasma flows, but from our experimental results we initially concluded we were unsuccessful, as the external field was frozen-out and prevented reconnection.
Our simulations, however, suggest that reconnection does occur later in the current pulse---if a higher peak driving current is used---and observing this is the goal of a future experimental campaign.
Additionally, simulations with a $\SI{1}{\tesla}$ external field suggest we could observe the delayed formation of a reconnection layer at the same peak current used in our experiments.
After these experiments, we developed an alternative technique using tilted wire arrays to study guide field reconnection \cite{Varnish2025}.

% We believe that the results presented in this paper may be relevant to the onset problem.
% Generically, our experiment has $\beta\sim 1$ plasma with anti-parallel magnetic fields, separated by a $\beta\ll1$ region with an ambient (or external) magnetic field which is perpendicular to the reconnecting fields.
% Such a configuration could occur, for example, in the convective region of the sun, where turbulence creates regions of varying density and magnetic field.
% The ambient magnetic field in the low density plasma prevents reconnection, allowing magnetic energy to be stored in the anti-parallel field components.
% When reconnection finally occurs, there is significant stored energy and reconnection will occur very rapidly.
% This is a model we intend to develop in the future with better-diagnosed experiments and higher-resolution numerical simulations.

\section{Acknowledgements}
This work was funded by NSF and NNSA under grant no. PHY2108050, and also supported by the NNSA Stewardship Science Academic Programs under DOE Cooperative Agreement DE-NA0004148.
TWOV acknowledges support from the MIT MathWorks fellowship. 
We would also like to thank Prof. Nuno F. Loureiro for his helpful comments on this paper; he will be dearly missed.
We are grateful to Prof. Jeremy Chittenden and Dr. Nikita Chaturvedi for their help with the GORGON simulations.
Additional thanks go to Simran Chowdhry, for her advice and discussions about the X-point collapse process.
We would also like to thank the engineering team at the MIT Plasma Science and Fusion Center (PSFC) for their work machining the load hardware.
We also appreciate the support of Dylan K. Robinson (MIT) in processing some of the interferograms analysed as part of this project.

\section{Declaration of Conflicts of Interest}

The authors have no conflicts of interest to disclose.

\section{Data Availability}

The data supporting the findings of this study are available from the corresponding author upon reasonable request.

% \bibliography{library,references}
% \bibliography{references_combined}
\bibliography{references_final}

\end{document}